\documentclass[onecolumn]{elsart3p}

\usepackage{graphicx}
\usepackage{amssymb}
\usepackage{amsmath}
\usepackage{color}
\usepackage[english,francais]{babel}

\newtheorem{e-proposition}[theorem]{Proposition}

\newtheorem{e-definition}[theorem]{Definition\rm}

\renewcommand{\theequation}{\arabic{equation}}
\def\og{\leavevmode\raise.3ex\hbox{$\scriptscriptstyle\langle\!\langle$~}}
\def\fg{\leavevmode\raise.3ex\hbox{~$\!\scriptscriptstyle\,\rangle\!\rangle$}}

\journal{CR Physique}

\begin{document}
%  You can place here the title of the dossier, if you know it,
%     firstly in English, then in French
\centerline{Title of the dossier/Titre du dossier}
\begin{frontmatter}


% \author{Name\thanksref{label2}}
\selectlanguage{english}
\title{
Electromagnetic and thermal responses 
in topological matter: \\
topological terms,
quantum anomalies and D-branes
}

% If all authors are at the same address, the [label1] can be suppressed

\selectlanguage{english}

\author[riken]{Akira Furusaki},
%\ead{author.name1@email.address1}
\author[cerg,cmrg,u-tokyo]{Naoto Nagaosa}
%\ead{author.name2@email.address2}
\author[tohoku]{Kentaro Nomura},
%\ead{author.name2@email.address2}
\author[uiuc]{Shinsei Ryu},
%\ead{author.name2@email.address2}
\author[yukawa,ipmu]{Tadashi Takayanagi}
%\ead{author.name2@email.address2}
\address[riken]{Condensed Matter Theory Laboratory, RIKEN, Wako, Saitama 351-0198, Japan}
\address[cerg]{Correlated Electron Research Group (CERG), RIKEN-ASI, Wako 351-0198, Japan}            
\address[cmrg]{Cross-Correlated Material Research Group (CMRG), RIKEN-ASI, Wako 351-0198, Japan}            
\address[u-tokyo]{Department of Applied Physics, The University of Tokyo, Hongo, Bunkyo-ku, Tokyo 113-8656, Japan}
\address[tohoku]{
 Institute for Materials Research, Tohoku University, Sendai 980-8577, Japan
}
\address[uiuc]{
 Department of Physics, University of Illinois,
 1110 West Green St, Urbana IL 61801
}
\address[yukawa]{
Yukawa Institute for Theoretical Physics,
Kyoto University, 
Kitashirakawa Oiwakecho, Sakyo-ku, Kyoto 606-8502, Japan
}
\address[ipmu]{
 Institute for the Physics and Mathematics of the Universe (IPMU),
 University of Tokyo, Kashiwa, Chiba 277-8582, Japan
}

\begin{abstract}
We discuss the thermal (or gravitational) responses
in topological superconductors
and in topological phases in general.
Such thermal responses (as well as electromagnetic responses
for conserved charge)
provide a definition of topological insulators and superconductors beyond the
single-particle picture.
In two-dimensional topological phases,
the Str\v{e}da formula for the electric Hall conductivity
is generalized to the thermal Hall conductivity.
Applying this formula to the Majorana surface states of three-dimensional
topological superconductors predicts cross-correlated responses between
the angular momentum and thermal polarization (entropy polarization).
We also discuss a use of D-branes in string theory as
a systematic tool to derive all such topological terms and
topological responses.
In particular, we relate the $\mathbb{Z}_2$ index of
topological insulators introduced by Kane and Mele
(and its generalization
to other symmetry classes and to arbitrary dimensions)
to
the K-theory charge of non-BPS D-branes,
and vice versa. 
We thus establish a link between
the stability of non-BPS D-branes
and the topological stability of 
topological insulators.

\vskip 0.5\baselineskip

\selectlanguage{francais}
\noindent{\bf R\'esum\'e}
\vskip 0.5\baselineskip
\noindent
{\bf Here is the French title. }
Your r\'esum\'e in French here.

%Now keywords/mots-clés
 \vskip 0.5\baselineskip
\noindent{\small{\it Mots-cl\'es~:} Mot-cl\'e1~; Mot-cl\'e2~;
Mot-cl\'e3}}
\end{abstract}
\end{frontmatter}

\selectlanguage{english}
% main text
\section{Introduction}
\label{}
% etc, etc

% \appendix

% \section{}
% \label{}

% The Acknowledgements are also a un-numbered section
%\section*{Acknowledgements}
% Acknowledgements text here

A vital role of topology in quantum transport phenomena in solids
has been recognized as a driving force of a {\it dissipationless} current.
The Berry curvature~\cite{Berry} in momentum and real spaces 
induces the velocity of Bloch electrons~\cite{Luttinger1954}, 
resulting in various effects such as the anomalous Hall effect~\cite{AHE} 
and the spin Hall effect~\cite{SHE}.
Even the heat current can be induced by the 
Berry curvature as observed in 
the anomalous Nernst effect~\cite{ANE} and 
the thermal Hall effect~\cite{Onose}. 
These quantum transport phenomena of topological origin
(or: {\it topological currents})
are regarded as a promising 
candidate for spintronics with low energy cost. 

In metallic systems the usual transport currents with dissipation
are dominant over these topological currents, and hence it is rather
difficult to identify the latter.  
A topological current, however, can be non-vanishing 
even in insulating systems 
--  
such systems on general ground are called topological insulator. 
This is so since it is the topological properties of electronic 
wavefunctions,
rather than a band structure of energy levels in solids, 
that are responsible for generation of a topological current.

The quantum Hall effect (QHE)
is a canonical example of topological insulators 
characterized by a topological (i.e., {\it quantized}) 
transport law
\cite{QHE_review,Thouless82,Kohmoto}.
Recently topological insulators
realized by strong spin-orbit interactions
in two and three dimensions
have been discovered
\cite{review_TIa,review_TIb,KaneMele,Roy,moore07,Fu06_3Da,bernevig06,Fu06_3Db}.
In particular, the three-dimensional (3d) topological insulator
is 
characterized by
the topological magnetoelectric effect 
\cite{Qi_Taylor_Zhang,Essin08} 
(the axion electrodynamics \cite{wilczekaxion}).
(See below for more details.)
As a consequence of non-trivial electrical wavefunctions in the bulk,
these topological insulators support anomalous boundary 
(edge or surface) modes,
whose gapless nature is topologically protected.

Analogously,
for superconductors (SCs) and superfluids,
one can consider
topological properties associated with the wavefunctions 
of fermionic quasiparticles;
Within the BCS mean-field theory,
the BCS (Bogoliubov-de Genne) Hamiltonian
is a fermion bilinear in the Nambu spinor.
The case where we have a quasi-particle
gap everywhere in momentum space
is an analogue of the band insulator. 
A topological SC is a SC with a full gap
and topologically nontrivial quasiparticle wavefunctions.
Canonical examples of topological SCs include, e.g.,
the 2d chiral $p$-wave SC
\cite{Read00,Maeno2003}.

In three dimensions,
the B phase of superfluid $^3$He was
recently identified as a new topological SC (superfluid)
\cite{Schnyder08,Roy08,Qi08b,footnote1,TSF}.
A recent surface transverse acoustic impedance measurement
reported in Refs.\ \cite{3HeB,Murakawa2010}
revealed a signature of
the surface Majorana fermion mode
on the surface of $^{3}\mathrm{He}$-B.
A copper-doped topological insulator 
(Cu$_x$Bi$_2$Se$_3$)
has been discussed
as a candidate of a 3d topological SC
\cite{Hor2010,FuBerg2010,Sasaki2011}.

The purpose of this article is to describe the response theory
of topological phases with a special focus on their thermal,
rather than electrical, transport.
A motivation for the thermal response is that it is well-defined
even for phases 
in which the electrical charge is not conserved,
such as topological SCs, or topological phases in spin systems.
While topological SCs (as well as topological insulators)
can be defined in terms of a topological invariant built out of,
within the BCS mean-field theory, quasiparticle wavefunctions,
the response theory gives a physically measurable
definition of topological phases,
which can be largely insensitive to microscopic details
-- a lesson we have learned from the physics of
the QHE, or the topological magnetoelectric effect
in the 3d topological insulator.
In particular, formulating the thermal response 
as a response to external gravitational field,
we will derive a thermal analogue of 
the topological magnetoelectric effect,
which uncovers an interesting
cross-correlation between 
thermal and 
mechanical responses,
in terms of the temperature gradient,
and an applied angular velocity, respectively
(Sec.\ \ref{gravitational response in topological superconductors}).

For the bulk of this article, we will focus on topological SCs
in two and three dimensions,
such as the 2d chiral $p$-wave SC or $^3$He B. 
However, we will also briefly describe
how such response theory can be systematically
constructed for a wider class of topological insulators
and SCs in the ``periodic table''
\cite{Schnyder08,KitaevLandau100Proceedings,Ryu_NJP}
(Sec.\ \ref{anomaly ladder and D-branes}).
In fact, we will illustrate that responses in topological phases
are closely related to quantum anomalies in field theories.
Such characterization of
topological phases in terms of anomalies
is expected to incorporate arbitrary strong interactions
as far as a bulk gap is not destroyed.

Interestingly,
such considerations lead to a natural link to
topological objects in string theory -- D-branes.
We will demonstrate, by considering a particular configuration
of D-branes, we can realize a field theory model of topological insulators
and superconductors with desired discrete symmetries
(Sec.\ \ref{anomaly ladder and D-branes}). 
The stability criterion of D-branes against Tachyon condensation 
is in one-to-one correspondence with the classification of 
topological insulators and superconductors (i.e., the periodic table
\cite{Schnyder08,KitaevLandau100Proceedings,Ryu_NJP}).
From condensed matter point of view,
the D-brane construction can be thought of as a
convenient tool that bridges K-theory classification
of fermionic Hamiltonians,
and liner response theory.

\section{electromagnetic response in the QHE and 3d topological insulator}

\subsection{the QHE in two dimensions}

Let us start this review by first illustrating the usefulness of the
effective field theory of 
linear response, by taking the QHE as an example.
The electromagnetic response
of the quantum Hall fluid is described by
the effective Chern-Simons action 
\begin{align}
I_{\mathrm{eff}}
=
\frac{e^2 k}{4\pi \hbar}
\int dt d^2x \,
\epsilon^{\mu\nu\lambda}
A_{\mu} \partial_{\nu} A_{\lambda},
\quad
k \in \mathbb{Z},
\end{align}
where $A_{\mu}$ is an external
electromagnetic gauge field.
The Chern-Simons action can be derived
by considering a coupling of
the (topological) insulator in question to
the external (background) electromagnetic field $A_{\mu}$,
and then by integrating out fermions to derive the effective
action for  $A_{\mu}$.

The Chern-Simons action encodes all types of
topological responses in the QHE.
(i)
The current
$\delta j^k(t,x)$
induced by $A_{\mu}$ is computed,
for time-independent vector potential, 
as
\begin{align}
%\delta \boldsymbol{j}(t,x)
%&=
\delta j^k(t,x)
=
\frac{\delta I_{\mathrm{eff}}}{\delta A_k(t,x)}
=
\frac{e^2 k}{2\pi\hbar}
\epsilon^{ki0}
\partial_i A_0.
\end{align}
This is the QHE,
$J^x=\sigma_{H} E^y$,
with the Hall conductivity
$\sigma_{H}= k e^2/(2\pi \hbar)$.
(ii) Similarly, one can compute the charge
$\delta \rho(t,x)$
induced by $A_{\mu}$ as:
\begin{align}
c \delta \rho(t,x)
&=
\delta j^0(t,x)
=
\frac{\delta I_{\mathrm{eff}}}{\delta A_0(t,x)}
=
\frac{e^2 k}{2\pi \hbar}
\epsilon^{0ij}
\partial_i A_j,
\end{align}
where $c$ is the speed of light. 
This means that
a (solitonic) flux tube binds $k$ units of charge,
which is, in cylinder geometry,
nothing but charge-pumping in Laughlin's thought experiment.
For uniform magnetic field $B^z$, the change in total electron charge is
\begin{align}
c \delta Q_e
=
%\frac{e^2 k}{2\pi\hbar}
\sigma_{H}
\delta B^z
\quad
 \Rightarrow
\quad
c
\frac{\delta Q_e}{\delta B^z}
=
c e \frac{\delta N_e}{\delta B^z}
=
\sigma_{H},
%\frac{e^2 }{2\pi \hbar} k
\label{EM streda 1}
\end{align}
where $N_e$ is the total electron number.
This is nothing but the Str\v{e}da formula
\cite{Streda1982,Smrcka1977}.
(iii)
The Chern-Simons term can be written,
for static configurations of $A_{\mu}$,
as
$
I_{\mathrm{eff}}
% &=
% \frac{e^2N}{4\pi}
% \int dt d^2x
% \left[
% 2 \epsilon^{0ij} A_0 \partial_i A_j
% +
% \epsilon^{i0j} A_i \partial_0 A_j
% \right]
% \nonumber \\%%%%%
=
%(e^2k)/(2\pi \hbar)
\sigma_H
\int dt d^2x\,
 \epsilon^{0ij} A_0 \partial_i A_j
=
%(e^2k)/(2\pi\hbar)
(\sigma_H/c)
\int dt d^2x\,
\phi B^z
$,
where
$A_0=:\phi/c$. 
If we define (the change in) magnetization by
\begin{align}
\delta M^z = \frac{\delta I_{\mathrm{eff}}}{\delta B^z}
=
(\sigma_{H}/c)
%\frac{e^2 k}{2\pi \hbar}
\delta \phi
\quad
\Rightarrow
\quad
ec\frac{\delta M^z}{\delta \mu}
=
\sigma_{H},
%\frac{e^2 k}{2\pi \hbar},
\label{EM streda 2}
\end{align}
where
$\mu =e\phi$ is the chemical potential
[$\mu N_e = (\mu/e) e N_e = \phi Q_e$].

\subsection{topological insulator in three dimensions}

Similarly, for the 3d topological insulator,
the electromagnetic response is encoded in the effective action
\cite{Qi_Taylor_Zhang,Essin08,wilczekaxion}
\begin{align}
I_{\mathrm{eff}}
&=
\frac{\theta e^2}{32\pi^2 \hbar c}
\int dt d^3 x\,
\epsilon^{\mu\nu\kappa\lambda}F_{\mu\nu}F_{\kappa\lambda}
=
\frac{\theta e^2}{4\pi^2 \hbar c}
\int dt d^3 x\,
\vec{E}\cdot \vec{B}.
\end{align}
This ``axion'' term can be derived,
similarly to the Chern-Simons action,
by integrating out fermions in the presence of the background
electromagnetic field.
The Dirac quantization condition
and time-reversal symmetry (TRS) restrict
$\theta$ to be quantized,
$\theta = 0, \pi$ (mod $2\pi$);
$\theta = 0$ for trivial insulators 
or vacuum
whereas
$\theta = \pi$ inside topological insulators.

As inferred from
the axion term,
the topological insulator features
topological magnetoelectric (ME) effect
\begin{align}
\vec{M}
&=
\frac{\delta I_{\mathrm{eff}}}{\delta \vec{B}}
=
\frac{\theta}{\pi}\frac{e^2}{2hc} \vec{E}
=
\frac{\theta}{\pi}
\frac{\alpha}{4\pi}
\vec{E},
\label{top ME 1}
\\%%%%%
   \vec{P}
&=\frac{\delta I_{\mathrm{eff}}}{\delta \vec{E}}
=\frac{\theta}{\pi}\frac{e^2}{2hc} \vec{B}
=\frac{\theta}{\pi}
\frac{\alpha}{4\pi}
\vec{B},
\label{top ME 2}
\end{align}
or,
quantized
electromagnetic poloarizability,
\begin{align}
\frac{\delta P_i}{\delta B_j}
=
\frac{\delta M_i}{\delta E_j}
=
\delta_{ij}
\frac{\theta}{\pi}
\frac{\alpha}{4\pi},
\end{align}
where
$\alpha= e^2/(\hbar c)$
is the fine structure constant.
Equations
(\ref{top ME 2})
and
(\ref{top ME 1})
are the 3d analogue of
Eqs.\ (\ref{EM streda 1})
and (\ref{EM streda 2}),
respectively.

\begin{figure}[tb]
\begin{center}
    \includegraphics[height=.35\textwidth]{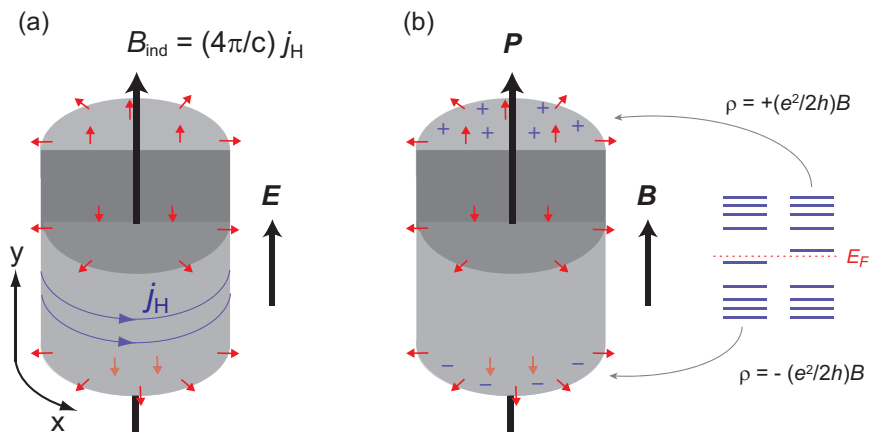}
\caption{
The three-dimensional topological insulator in cylindrical geometry;
(a) the induced magnetization follows the direction of the applied
electric field;
(b) the induced polarization follows the direction of the applied
magnetic field. 
In both cases, time-reversal symmetry is broken on the surface,
by putting, say, magnetic impurities, 
as indicated by red arrows. 
In (a), the magnetization is caused by the surface Hall current
$\vec{j}_H$.
In (b), the polarization is caused by an excess or a deficit charge
on the top or bottom surface
[see Eq.\ (\ref{EM streda 1})]. 
}
\label{fig1}
\end{center}
\end{figure}

The magnetization $\vec{M}$ in Eq.\ (\ref{top ME 1}),
which follows the direction of the external
electric field $\vec{E}$,
is generated by
the surface QHE 
(Fig.\ \ref{fig1}a);
when a topological insulator is 
in contact with
a topologically trivial insulator (or simply vacuum),
the $\theta$-angle in the axion term jumps by $\pi$
at the interface.
As the axion term is the total derivative
of the Chern-Simons term, such interface
(where TRS is weakly broken)
is accompanied by the half quantized surface Hall current
with $\sigma_{H}=\pm e^2/(2h)$
which generates a bulk magnetization $\vec{M}$.
Similarly,
when an external magnetic field $\vec{B}$ is applied,
according to Eq.\ (\ref{EM streda 1}),
it induces an excess or a deficit of charge
on
the surfaces which are orthogonal to $\vec{B}$ --
this is the source of a bulk electric polarization $\vec{P}$
in Eq.\ (\ref{top ME 2})
(Fig.\ \ref{fig1}b).

\section{gravitational response in topological superconductors}
\label{gravitational response in topological superconductors}

\subsection{gravitoelectromagnetism}

\paragraph{energy-gravity coupling}

We now develop the thermal response theory of topological phases,
with an eye, in particular, on applications to topological SCs.
Our strategy here is, based on Luttinger's idea
\cite{Luttinger},
to follow as close as possible the linear response theory
for the electric charge:
When we study charge response,
we consider a small external probe field for the charge density,
$H_I \sim \int d^d x\, \phi(\vec{x}) \rho(\vec{x})$.
The same strategy can be adopted since energy is conserved;
as in the charge response, we can consider an external (fictitious)
source term which couples to the energy (Hamiltonian) density 
$\varepsilon(\vec{x})$,
$H_I \sim \int d^d x\, \phi_g (\vec{x}) 
\varepsilon(\vec{x})/v^2$,
where
we have introduced $v$ which has the dimension of velocity
to assign a proper dimension 
to $\phi_g$ (see below).

The external source $\phi_g$ can be thought of as a
fictitious gravitational field in the presence of the Lorentz invariance,
as it allows us to identify the energy as the mass,
$\varepsilon(\vec{x})= m(\vec{x}) v^2$,
where $m(\vec{x})$ is the mass density.
For the gravitational theory of nature
(not for Luttinger's fictitious gravity that is a mere device to develop
a linear response theory),
$v$ should be replaced by the speed of light $c$,
whereas in condensed matter systems with emergent Lorentz symmetry,
$v$ is the ``Fermi'' velocity (see below for more discussion).
The coupling of the system to the potential $\phi_g$
can then be written as
$H_I \sim \int d^d x\, \phi_g (\vec{x}) m(\vec{x})$,
where, 
from the analogy to the electromagnetism,
mass can be thought of as a ``charge'' coupled to gravitational field.

With Lorentz invariance,
in more covariant language (in the Lagrangian language)
the gravitational coupling with the (thermal) energy current is introduced as follows:
The energy density $\varepsilon$ 
and the energy current $\vec{j}_E$
are components of the energy-momentum tensor
$T^{\mu\nu}$,
$\varepsilon=T^{00}$ and $j^i_E=v T^{0i}$.
They are thus coupled to 
the variation of spacetime metric $g_{\mu\nu}$ as
$  
-(1/2)\int dt d^dx\,  \sqrt{-g}\, T^{\mu\nu}\delta g_{\mu\nu}
$
in the Lagrangian, 
where $g= \mathrm{det}\, g_{\mu\nu}$
--
in a way analogous to the way 
the charge current $j^{\mu}$
couples with the external electromagnetic potential
as $-\int dt d^d x\, j^{\mu}A_{\mu}$.

\paragraph{gravitoelectric field}

A spatial gradient in energy density
inevitably causes a temperature gradient,
as one can infer from the thermodynamic
equality
$dU = T dS$
as follows.
(Here,
$U$ is the internal energy,
$S$ the entropy,
and
$T$ the temperature).
Let us divide the total system into two
subsystems (subsystems 1 and 2).
Equilibrium between the two is achieved when the total entropy is maximized
$dS = dS_1 + dS_2 = 0$.
Since energy is conserved, $dE_2 = -dE_1$, 
and hence
$dS_1/dE_1 -
dS_2/dE_2 = 0$.
I.e, $T_1 = T_2$.
Let us now turn on a gradient in the gravitational potential, 
so that the gravitational potential felt by Subsystem 1 and 2
differs by $\delta \phi_g$.
In this case, 
we have
$dE_2 = -dE_1(1 + \delta \phi_g /v^2)$.
This suggests the generation of a temperature difference 
$T_2 = T_1 (1 + \delta \phi_g /v^2)$.
In other words, 
we can view 
the ``electric'' field $\vec{E}_g$
associated with the gradient of $\phi_g$, 
which we call
``gravitoelectric field'',
as temperature gradient,
\begin{align}
\vec{E}_g :=
- \vec{\nabla} \phi_g
=
v^2 T^{-1} \vec{\nabla} T. 
\end{align}

\paragraph{gravitomagnetic field}

The analogy with electromagnetism
can be further put forward --
such formalism is called gravitoelectromagnetism
\cite{GEM}.
For our purposes to develop a linear response theory
for thermal transport,
we put an external gravity field which is infinitesimally
small. We can thus write the metric as
$
g_{\mu\nu} = \eta_{\mu\nu} + h_{\mu\nu}
$
where $\eta_{\mu\nu}$ is the metric of flat space-time.
In the presence of matter,
in the post Newtonian limit,
we keep only 
$h_{00}$, $h_{ii}$, and $h_{0i}$
[which are of order $\mathcal{O}(v^{-2})$], 
whereas
$h_{ij}=\mathcal{O}(v^{-4})$ ($i\neq j$) are neglected
\cite{GEM}:
\begin{align}
ds^2 =
v^2 (-1 
+
 2 \phi_g/v^2 )dt^2
- \frac{4}{v} \vec{A}_g \cdot d\vec{x} dt
+ (1 
+
 2 \phi_g/v^2 )
d\vec{x}\cdot d\vec{x},
\label{metric, GEM}
\end{align}
where we have introduced
the gravitomagnetic potential,
$A_{g,i}:= -(v^2/2) h_{0i}$. 
For small $h_{\mu\nu}$,
the Einstein equation can be linearized 
up to these non-zero components of $h_{\mu\nu}$ kept in Eq.\ (\ref{metric, GEM}),
and the resulting equation looks structurally identical
to the Maxwell equation.
(However, note that we are considering
an external gravity field as a source which does not have its own dynamics).
As we have seen,
the gravitoelectric field $\vec{E}_g$
corresponds to temperature gradient.
What does the gravitational analogue of magnetic field,
the gravitomagnetic field, $\vec{B}_g$,
correspond to?
It turns out that
$\vec{B}_g$ can be understood as the angular velocity vector
of rotating systems,
\begin{align}
\vec{B}_g:= \vec{\nabla}\times \vec{A}_g = 
v \vec{\Omega}.
\end{align}
For a
system rotating with the frequency
$\vec{\Omega}=\Omega^z \vec{z}$,
this can be understood by making a coordinate transformation
from the rest frame to the rotating frame
in which the metric in the
polar coordinates
$(t,r,\varphi,z)$
takes the form
$
ds^2\simeq
-v^2dt^2+2 \Omega^z r^2d\varphi dt +r^2d\varphi^2+dr^2+dz^2.
$
One can then read off, from the definition of the
gravitoelectromagnetic field,
the non-zero gravito gauge potential
$A_g^{\varphi} = (v/2) \Omega^z r$.
In Cartesian coordinates,
$ \vec{A}_g=
(v/2) \Omega^z \hat{z}\times \vec{x},
$
and
$\vec{B}_g=
\vec{\nabla} \times\vec{A}_g=
v 
\Omega^z \hat{z}
$.

\subsection{thermal Str\v{e}da formula for 2d topological SCs}

We now use the formalism described above to study
thermal transport of topological SCs in two and three dimensions.
The goal here is to establish
a gravitational analogue of
the Str\v{e}da formula
(\ref{EM streda 1}-\ref{EM streda 2})
and
the topological ME effect
(\ref{top ME 1}-\ref{top ME 2})
\cite{NomuraStreda2011}.

Let us start with
2d topological fluid which does not have time-reversal symmetry.
To draw a parallelism with the charge response,
we introduce
the moment of the thermal current $\vec{M}_T$
as
$\vec{M}_T=\vec{M}_E-(\mu/e)\vec{M}$, where
$
M^{\mu\nu}_E=
\left\langle x^{\mu}T^{0\nu}-x^{\nu}T^{0\mu}\right\rangle,
$
and $M^z_E=M^{12}_E$.
(Similarly,
the thermal current can be decomposed into
$\vec{j}_T=\vec{j}_E-(\mu/e)\vec{j}$, where
$\vec{j}$ is the electric charge current.)
Here the average has to be taken at finite temperature:
$\langle \cdots \rangle\equiv
\sum_n f(\varepsilon_n)\langle n|\cdots|n\rangle$
where $\varepsilon_n$ and $|n\rangle$ are eigenvalue and eigenstate
of the Hamiltonian $\mathcal{H}$,
and $f(\varepsilon_n)$ is the Fermi distribution function.%
\footnote{
Below, we will consider the part of the thermal current
carried by $\vec{j}_E$ (and hence $\vec{M}_E$)
specializing to the case of $\mu =0$.
This is valid in discussing topological SCs.}
\footnote{
Our convention for $\phi_g$, $\vec{A}_g$, and $\vec{M}_E$
in this review differs from the one employed 
in Ref.\ \cite{NomuraStreda2011};
there is a factor of two difference 
in defining $\phi_g$ and $\vec{A}_g$
[see Eq.\ (\ref{metric, GEM})].
Accordingly, there is a similar 
factor of 1/2 difference in defining $\vec{M}_E$.
These factors are due to the spin 2 nature of gravitons, 
and are convenient since the resulting 
linearized Einstein equation does not 
have such factors of two, and the similarity
with the Maxwell equation is clearer. 
Also, in the definition of $\phi_g$, 
we have put an extra factor of $v^2$. 
}
We can then
obtain
\begin{align}
 \kappa_{H}^{}=
\frac{v}{2} \frac{\partial M^z_T}{\partial T}
\label{streda-je}
\end{align}
from the Kubo formula
for the thermal Hall conductivity\footnote{
While 
in the bulk, in the absence of interactions,
the Kubo formula gives us one of the most direct 
ways to get 
the Str\v{e}da formula for $\sigma_{H}$ and $\kappa_H$,
the applicability of the Str\v{e}da formulas is not limited
to non-interacting systems, as they 
can be also derived assuming the presence of chiral edge states.
See Ref.\ \cite{NomuraStreda2011} for details. 
\label{footnoteCFT}
}
\cite{NomuraStreda2011}.

To see the physical meaning of $\vec{M}_E$,
note that the definition of $M_E^{\mu\nu}$
is similar to the orbital angular momentum:
$
 L^{\mu\nu}=
(1/c)   \left\langle x^{\mu}T^{\nu0}-x^{\nu}T^{\mu0}\right\rangle.
$
Indeed,
when there is a relativistic invariance,
the energy-momentum tensor can be symmetrized so that $T^{\mu\nu}=T^{\nu\mu}$,
and thus $M^{\mu\nu}_E= cL^{\mu\nu}$.
While this is not the case in condensed-matter systems in general,
the (pseudo) Lorentz invariance can emerge in solids
at low energies, such as in graphene or in topological superconductors
where electrons or quasiparticles obey the Dirac or Majorana equation.
In these systems the Fermi velocity $v$
plays a role of $c$,
and 
the $M^{\mu\nu}_E$ tensor is related to the orbital angular momentum
as $\vec{M}_E= v \vec{L}$.
Thus, 
$ \kappa_{H}^{}=
(v^2/2) 
(\partial L^z/\partial T).
$
This is an analogue of 
the electromagnetic Str\v{e}da formula (\ref{EM streda 2}). 

To derive ``the other half'' of the Str\v{e}da formula
[a thermal analogue of Eq.\ (\ref{EM streda 1})], 
we note that the variation of free energy is given by
$
dF=
-SdT-\vec{L}\cdot d\vec{\Omega}
$
where $S$ is the entropy. 
In terms of the gravitoelectric and 
gravitomagnetic fields,
this can be written as 
$
dF=
-(TS/v^2) (v^2 T^{-1}dT)-
(\vec{L}/v)\cdot d(v\vec{\Omega})
=
-(Q_T /v^2) (d \phi_g)-
(\vec{M}_E/v^2) \cdot d\vec{B}_g
$,
where 
$Q_T=TS$ is the thermal energy density,
which couples to $\phi_g/v^2$.
From the Maxwell relation, 
we thus obtain
\begin{align}
\kappa_{H}^{}
=
\frac{v^3}{2T}\left(\frac{\partial M^z_E}{\partial\phi_g}\right)_{B^z_g}
=
\frac{v^3}{2T}\left(\frac{\partial Q_T}{\partial B_g^z}\right)_{\phi_g}. 
\label{streda-je2}
\end{align}
This is
the thermal analogue of
the Str\v{e}da formula for the charge Hall conductivity,
in that $Q_T$ is the zeroth component of the energy current
as $eN_e$ is in the charge current.
With the pseudo-relativistic invariance,
this can be written as
\begin{align}
\kappa_{H}^{}
=\frac{v^2}{2}
\left(\frac{\partial L^z}{\partial T} \right)_{\Omega^z}
=\frac{v^2}{2}
\left(\frac{\partial S}{\partial \Omega^z}\right)_{T}.
\label{streda-je3}
\end{align}

\subsection{cross-correlated response of 3d topological SC}
\label{cross-correlated response of 3d topological SC}

The thermal Str\v{e}da formula derived above for 2d topological
fluid can be used to study the response of
3d topological SCs to the temperature gradient and the rotation; 
Given that the fermionic quasiparticles are fully gapped in the bulk,
all (topological) transport phenomena,
in the presence of a boundary (surface), 
can essentially be discussed by looking at 
the surface transport.  
For simplicity, 
one can consider a cylindrical geometry (as in Fig.\ \ref{fig1}),
apply the temperature gradient 
$\partial_zT$
or external rotation
$\Omega^z$,
and discuss the responses, which are mediated by the surface. 
(As in the electromagnetic responses, 
we weakly break TRS at the surface and hence the surface is gapped).  
Such thought experiments lead to,
for the (induced) thermal polarization $\vec{P}_E$ defined by
$\delta Q_T=-\vec{\nabla} \cdot \vec{P}_E$,
and 
for the moment of the energy current $\vec{M}_E$,
\begin{align}
\vec{M}_E &= (T\kappa_{H}/v^{3}) \vec{E}_g,
\label{grav top ME a}
 \\%%%%%
\vec{P}_E &= (T\kappa_{H}/v^{3}) \vec{B}_g,
\label{grav top ME b}
 \\%%%%%
 \frac{\delta M_{E,i}}{\delta E_{g,j}}
 & =
 \frac{\delta P_{E,i}}{\delta B_{g,j}}
=
\delta_{ij} 
\frac{\theta}{\pi} 
\frac{T\kappa_{H}}{v^{3}}. 
\label{grav top ME}
\end{align}

The parallel between
the electromagnetic and gravitational
cases are obvious.
Observe, however, that
the gravitational response is not 
quantized as in the charge response, because of the presence of
the velocity $v$.
The situation is somewhat similar to 
the detection of the conformal anomaly (central charge)
from specific heat 
in 1d quantum systems at criticality; 
while 
the central charge $C$ %$c_{c.c.}$ 
for a given quantum critical system in 1d
is a dimensionless universal parameter, 
it shows up in specific heat $C_V$
together with the velocity $v$
as $C_V = C \pi T /(3v)$.

\paragraph{possible experiments}

The thermal (gravitational) analogue of the 
Str\v{e}da formula 
[Eq.\ (\ref{streda-je3})]
and 
topological ME effect
[Eq.\ (\ref{grav top ME})]
can be tested experimentally.
An external angular velocity $\Omega^z$
results in the change in temperature (in 2d topological SCs)
and
thermal polarization (in 3d topological SCs).
If the heat capacity of the system is sufficiently small,
these may not be difficult to measure.

In three dimensions, 
the ``dual'' response, 
i.e., 
the response 
to the applied temperature gradient
[Eq.\ (\ref{grav top ME b})], 
can be detected by making use of the Einstein-de Hass effect.
Let us assume a cylindrical 3d topological SC is suspended by a thin string
and apply thermal gradient (as in Fig.\ \ref{fig1}a).
This induces surface energy current with angular momentum
$L^z$, according to Eq.\ (\ref{grav top ME b}).
By the conservation law of total angular momentum,
it must be compensated by a mechanical angular momentum of the material,
which can be directly measured in principle.

Note that in both responses to $\Omega^z$ and 
to temperature gradients, 
the part of the responses of our interest are contributions
from the fermionic quasiparticles. They should be distinguished 
from the contributions from bosonic excitations such as vortices. 
If 
$\Omega^z$ is larger than the critical angular velocity $\Omega_{c1}$
above which vortices are introduced in the bulk of the sample,
an extra contribution to 
thermal polarization would be generated.

\section{anomaly ladder and D-branes}
\label{anomaly ladder and D-branes}

\subsection{integrating out fermions and chiral anomaly}

We have constructed the response theory of 3d topological SCs
starting from the thermal Str\v{e}da formula of
2d topological SCs.
We now discuss the response of topological SCs (and insulators)
to external gravitational field
from a field theoretical point of view
\cite{Ryu2011,Wang2011}.
Our goal here is to derive a gravitational analogue
of the axion term in the electromagnetic response,
and show that it is related to quantum anomaly (chiral anomaly).
In this section, we will use natural units
$e=c=\hbar=1$, and set the Fermi velocity $v$ to be unity
for simplicity.

Let us work with an example;
A canonical example of the 3d
topological SC is
the B phase of $^3$He,
which is described, in momentum space,
by the following BdG Hamiltonian:
$H =
(1/2)
\int d^3 k\,
\Psi^{\dag}(\vec{k})
\mathcal{H}(\vec{k})
\Psi(\vec{k}),
$
where
$\Psi^{\dag}(\vec{k})
=
\big(
c^{\dag}_{\uparrow, \vec{k}},
c^{\dag}_{\downarrow, \vec{k}},
c^{\ }_{\uparrow,-\vec{k}},
c^{\ }_{\downarrow,- \vec{k}}
\big)
$
is the Nambu spinor
composed of fermionic creation/annihilation
operators ($c^{\dag}_{s, \vec{k}}/c^{\ }_{s, \vec{k}}$)
of a $^{3}$He atom
with spin $s$ and momentum $\vec{k}$,
and the kernel
$\mathcal{H}(\vec{k})$
takes the following form:
\begin{align}
\mathcal{H}(\vec{k})
=
\left(
\begin{array}{cc}
\xi(\vec{k}) & \Delta(\vec{k}) \\
% -\Delta^{*}
\Delta^{\dag} (\vec{k})
& -\xi(-\vec{k})
\end{array}
\right),
\quad
\xi(\vec{k})
=\frac{k^2}{2M} - \mu,
\quad
\Delta(\vec{k})
=
|\Delta| \vec{k}
%\boldsymbol{d}(\vec{k})
\cdot
\vec{s}
({i}s_y),
\label{He3}
\end{align}
where $M$ is the mass of a $^{3}$He atom,
$\mu$ is the chemical potential,
and $|\Delta|$ is the amplitude of the pair potential.
With the $\vec{d}$-vector pointing parallel
to momentum,
$\vec{d}(\vec{k})=|\Delta|\vec{k}$,
there is an isotropic gap everywhere on
the 3d fermi surface.
The critical point at $\mu=0$ separates
topologically trivial ($\mu<0$)
and non-trivial ($\mu>0$)
phases, which are characterized by
an integral topological invariant of symmetry class DIII (the winding number)
$\nu=0$ and $\nu=1$, respectively
\cite{Schnyder08}.
We will henceforth set $|\Delta|=1$
and drop the $\mathcal{O}(k^2)$ term
in $\xi(\vec{k})$,
$\xi(\vec{k})\to -\mu \equiv m$.
With a suitable unitary transformation,
the BdG Hamiltonian
is written in terms of
the $4\times 4$ Dirac matrices
$
\{\alpha_{i=1,2,3}, \beta\}
$
as
$\mathcal{H}
=
-{i} \sum_{i=1}^3 \alpha_i \partial_i 
+m \beta.
$
Upon this linearizeation,
$\mu$ can be thought of as a
``Dirac mass term''.

We introduce an external gravity field
which couples to the Dirac Hamiltonian (Lagrangian) as,
$S [\bar{\psi},\psi,e]
=
\int  dt d^3 x\,
\sqrt{-g}
\mathcal{L}$,
\begin{align}
\mathcal{L}
=
\bar{\psi}
e^{\ }_{a}{ }^{\mu}
{i}
\gamma^{a}
\Big(
\partial_{\mu}
-
\frac{{i}}{2}
\omega_{\mu}{ }^{cd}
\Sigma_{cd}
\Big)
\psi
-
m
\bar{\psi}
\psi,
\end{align}
where
$\mu,\nu,\ldots=0,1,2,3$
is the space-time index,
and
$a,b,\ldots=0,1,2,3$
is the flat index;
$e_{a}{ }^{\mu}$ is vielbein,
and $\omega_{\mu}{ }^{ab}$ is a spin connection;
$\Sigma_{ab}=\left[\gamma_a,\gamma_b\right]/(4 i)$.
The effective action $I_{\mathrm{eff}}[m,e]$
for the gravitational field is then obtained from
the fermionic path integral
$
I_{\mathrm{eff}}[m,e]
=
-{i} \ln \int\mathcal{D}[\bar{\psi},\psi]
\exp( {i} S[\bar{\psi},\psi,e])
$.

Below, we are interested in the topological term of
the effective action, which is, in the imaginary time path integral,
an imaginary part of the $S_{\mathrm{eff}}[m,e]$.
This part can be computed by making use of chiral anomaly as follows
\cite{Hosur2008}.
We first observe that the continuum
Hamiltonian $\mathcal{H}$
enjoys a continuous chiral symmetry:
we can flip the sign of mass, in a continuous fashion,
by the following chiral rotation
$
\psi\to\psi=e^{{i}\phi \gamma_{5}/2}\psi'
$,
$
\bar{\psi} \to \bar{\psi}=
\bar{\psi}' e^{{i}\phi\gamma_{5}/2},
%\psi^{\dag} \to \psi^{\dag}=
%\psi^{\dag \prime} e^{-{i}\phi\gamma_{5}/2},
%\label{chro}
$
under which the mass is rotated as
\begin{eqnarray}
m'(\phi)
=
me^{{i}\phi \gamma_{5}}=
m\left(\cos\phi+{i}\gamma_{5}\sin\phi \right),
\end{eqnarray}
so that $m'(\phi=0)=m$ and $m'(\phi=\pi)=-m$.
Since $m$ can continuously be
rotated into $-m$, one would think, naively,
$I_{\mathrm{eff}}[m]=I_{\mathrm{eff}}[-m]$.
This naive expectation is, however, not true because of chiral anomaly.
The chiral transformation which rotates $m$ continuously
costs the Jacobian $\mathcal{J}$ of the path integral measure,
\begin{eqnarray}
\mathcal{D}\left[\bar{\psi},\psi\right]
=
\mathcal{J}\mathcal{D}\left[\bar{\psi'},\psi'\right].
\end{eqnarray}
This chiral anomaly (the chiral Jacobian $\mathcal{J}$) is responsible
for the gravitational analogue of the axion term (the $\theta$-term).
The Jacobian $\mathcal{J}$ can be computed explicitly by
the Fujikawa method
\cite{fujikawa}
as
\begin{align}
\label{GravitationalInstantonTerm}
I^{\theta}_{\mathrm{eff}}
=
-\ln \mathcal{J}
=
%\frac{  \theta}{2\times 384\pi^2}
%\theta
%\left[
\frac{1}{2}
%\frac{ 1}{2\times 384\pi^2}
\frac{1}{2}\frac{\theta}{384\pi^2}
\int d^4 x
\sqrt{-g} \epsilon^{cdef}
R^{a}{ }_{b cd }
R^{b}{ }_{a ef}
%\right]
,
\end{align}
where
$R= d\omega + \omega \wedge \omega$
($R^{a}{ }_{b\mu\nu}=
\partial_{\mu}\omega_{\nu}^{a}{ }_b
-
\partial_{\nu}\omega_{\mu}^{a}{ }_b
+
[\omega_{\mu}, \omega_{\nu}]^a{ }_{b}
$)
is
the Riemann curvature tensor;
as in the electromagnetic response,
the $\theta$-angle
is fixed to either $\theta=0$ or $\theta=\pi$ by time-reversal symmetry.
The former corresponds
to a topologically trivial state, and
$\theta=\pi$ to a topologically non-trivial
state.
The theta term in the gravitational effective action
(\ref{GravitationalInstantonTerm})
(``the gravitational instanton term'')
is an analogue of the axion term 
$\propto \theta \vec{E}\cdot \vec{B}$
in the electromagnetic effective action;
There is an obvious structural parallelism between
the electromagnetic and gravitational cases
\cite{comment_grav_instanton}.

To make the connection with the existence of
topologically protected surface modes,
we note,
when there are boundaries (say) in the $x^3$-direction
at $x^3=L_+$ and at $x^3=L_-$,
the gravitational instanton term
$I^{\theta}_{\mathrm{eff}}$,
at the non-trivial time-reversal invariant
value $\theta=\pi$ of the angle $\theta$,
can be written
in terms of the gravitational Chern-Simons terms at the boundaries,
$I^{\theta}_{\mathrm{eff}}
=
I_{\mathrm{CS}}|_{x^3=L_+}
-I_{\mathrm{CS}}|_{x^3=L_-},
$
where ($i,j,k=0,1,2$)
\begin{align}
I_{\mathrm{CS}}
=
\frac{1}{2}
\
\frac{1}{4\pi}
\
\frac{ c'}{24}
\int d^3 x\,
\epsilon^{ijk}
\mathrm{tr}\,
\Big(
\omega_{i} \partial_{j} \omega_{k}
+
\frac{2}{3}
\omega_{i}
\omega_{j}
\omega_{k}
\Big),
\end{align}
with $c'=1/2$.
This value of the coefficient of the gravitational Chern-Simons term
is one-half of the canonical value
$
(1/4\pi)
\times
(c'/24)
$
with $c'=1/2$.
As discussed by
Volovik
\cite{Volovik90}
and
by
Read and Green
\cite{Read00}
in the context of the 2d chiral $p$-wave SC,
the coefficient of the gravitational Chern-Simons term
is directly related to the thermal Hall conductivity, which in our case
is carried by the topologically protected surface modes.

\subsection{anomaly ladder and periodic table}
\label{anomaly ladder and periodic table}

All types of responses we have discussed so far
(the QHE, the topological ME, and their gravitational (thermal) analogues)
are related to quantum anomaly;
the Chern-Simons terms in the QHE (both electromagnetic and gravitational)
is a manifestation of parity anomaly;
the axion term can be derived, as we have seen above, from chiral anomaly.
In fact, a wider class of topological
insulators and SCs in the periodic table
(Table \ref{periodic table with gravity})
can be related to quantum anomalies.

\begin{table}
\begin{center}
\begin{tabular}{cccccccccccccc}\hline\hline
Symmetry Class$\backslash d$ &
%Cartan$\backslash d$ &
\textcolor{magenta}{0}  &
\textcolor{green}{1} &
\textcolor{red}{2} &
\textcolor{blue}{3} &
\textcolor{magenta}{4} &
\textcolor{green}{5} &
\textcolor{red}{6} &
\textcolor{blue}{7} &
%\textcolor{magenta}{8} &
%\textcolor{green}{9} &
%\textcolor{red}{10} &
%\textcolor{blue}{11} &
$\cdots$ \\ \hline
A   & $\mathbb{Z}$& 0  & $\mathbb{Z}$& 0 &  $\mathbb{Z}$& 0  & $\mathbb{Z}$& 0
%&    $\mathbb{Z}$& 0 &  $\mathbb{Z}$& 0
&  $\cdots$ \\
AIII & 0& $\mathbb{Z}$& 0  & $\mathbb{Z}$& 0 &  $\mathbb{Z}$& 0  & $\mathbb{Z}$
%& 0 &  $\mathbb{Z}$& 0 &  $\mathbb{Z}$
&  $\cdots$ \\ \hline
AI  & \textcolor{magenta}{$\mathbb{Z}^{\spadesuit}$} & 0 & 0
    & 0 & $2\mathbb{Z}$ & 0
    & $\mathbb{Z}_2$ & $\mathbb{Z}_2$
%& \textcolor{magenta}{$\mathbb{Z}^{\spadesuit}$}    & 0 & 0 & 0
& $\cdots$ \\
BDI & $\mathbb{Z}_2$ &
\textcolor{green}{$\mathbb{Z}^{\clubsuit}$} & 0 & 0
    & 0 & $2\mathbb{Z}$ & 0
    & $\mathbb{Z}_2$
%& $\mathbb{Z}_2$ & \textcolor{green}{$\mathbb{Z}^{\clubsuit}$}    & 0 & 0
&  $\cdots$ \\
D   & $\mathbb{Z}_2$ & $\mathbb{Z}_2$ &
\textcolor{red}{$\mathbb{Z}^{\heartsuit}$}
    & 0 & 0 & 0
    & $2\mathbb{Z}$  & 0
%& $\mathbb{Z}_2$    & $\mathbb{Z}_2$ & \textcolor{red}{$\mathbb{Z}^{\heartsuit}$}  & 0
& $\cdots$ \\
DIII& 0 & $\mathbb{Z}_2$ & $\mathbb{Z}_2$ &
\textcolor{blue}{$\mathbb{Z}^{\diamondsuit}$ }
    & 0 & 0 & 0
    & $2\mathbb{Z}$
%& 0 & $\mathbb{Z}_2$     & $\mathbb{Z}_2$ & \textcolor{blue}{$\mathbb{Z}^{\diamondsuit}$}
&  $\cdots$ \\
AII & $2\mathbb{Z}$  & 0 & $\mathbb{Z}_2$
    & $\mathbb{Z}_2$ & \textcolor{magenta}{$\mathbb{Z}^{\spadesuit}$} & 0
    & 0 & 0
%& $2\mathbb{Z}$     & 0 & $\mathbb{Z}_2$ & $\mathbb{Z}_2$
&  $\cdots$\\
CII & 0 & $2\mathbb{Z}$  & 0 & $\mathbb{Z}_2$
    & $\mathbb{Z}_2$ &
\textcolor{green}{$\mathbb{Z}^{\clubsuit}$} & 0
    & 0
%& 0 & $2\mathbb{Z}$     & 0 & $\mathbb{Z}_2$
&  $\cdots$\\
C   & 0  & 0 & $2\mathbb{Z}$
    & 0 & $\mathbb{Z}_2$  & $\mathbb{Z}_2$
    & \textcolor{red}{$\mathbb{Z}^{\heartsuit}$} & 0
%& 0     & 0 & $2\mathbb{Z}$ & 0
& $\cdots$ \\
CI  & 0 & 0  & 0 & $2\mathbb{Z}$
    & 0 & $\mathbb{Z}_2$  & $\mathbb{Z}_2$
    & \textcolor{blue}{$\mathbb{Z}^{\diamondsuit}$}
%& 0 & 0     & 0 & $2\mathbb{Z}$
& $\cdots$ \\ \hline\hline
\end{tabular}
\end{center}
\caption{
Periodic table of topological insulators and superconductors
for the 10 symmetry classes in various spatial dimensions.
Topological insulators (superconductors)
in the complex symmetry classes (A and AIII) are related to the
chiral U(1) anomaly.
The primary series of the topological insulators (superconductors)
with an integer ($\mathbb{Z}$) classification
in the eight real symmetry classes are located on the diagonal
in the table.
In even space-time dimensions (odd space dimensions) they are predicted from
the chiral anomaly in the presence of background gravity
(\textcolor{blue}{$\mathbb{Z}^{\diamondsuit}$}),
and from the chiral anomaly in the presence of both background gravity and
U(1) gauge field
(\textcolor{green}{$\mathbb{Z}^{\clubsuit}$}).
The topological response of topological phases
in odd space-time dimensions
(\textcolor{red}{$\mathbb{Z}^{\heartsuit}$} and
\textcolor{magenta}{$\mathbb{Z}^{\spadesuit}$})
follows from 
their higher-dimensional ancestor 
(\textcolor{blue}{$\mathbb{Z}^{\diamondsuit}$} and
\textcolor{green}{$\mathbb{Z}^{\clubsuit}$}),
respectively.
}
\label{periodic table with gravity}
\end{table}

Let us consider topological insulators and SCs
with an integer topological invariant
located on the diagonal of the periodic table
(Table \ref{periodic table with gravity}),
which we call the ``primary series''.
(The $\mathbb{Z}_2$ topological phases which ``descend''
from the primary series can be called
1st and 2nd descendants \cite{Ryu_NJP}.)
For the primary series,
the effective action for electromagnetic and gravitational response
can be derived essentially by repeating the procedure discussed above;
taking a Dirac representative in $D=d+1=\mbox{even}$,
coupling it to the background electromagnetic and gravitational fields,
and then integrate out fermions;
chiral anomaly in $D=d+1=\mbox{even}$ dimensions allows 
us to calculate the imaginary
part of the action (i.e., the part which encodes topological part
of the response).
The result is summarized as follows:
\begin{align}
\delta \ln Z  =
2\pi {i}
\int_{M_D}
\left.
\mathrm{ch}(E) \hat{A}(R)
\right|_D.  \label{anomalyp}
\end{align}
$\delta \ln Z$
represents
the change in the effective action under
the infinitesimal chiral transformation;
by integrating it one gets the topological action
for the linear response.
$\mathrm{ch}(E)$ denotes the Chern character of the vector bundle $E$,
which is explicitly given by
$\mathrm{ch}(E)=\mathrm{tr}\,[e^{F/(2\pi)}]$
in terms of its field strength $F$.
$\hat{A}(R)$ is the A-roof genus and takes the form
\begin{equation}
\hat{A}(R)
=1+\frac{1}{192\pi^2}
\mathrm{tr}\, [R^2]+\cdots,
\end{equation}
where $R$ is the curvature two-form on the manifold $M_D$.
Since (\ref{anomalyp})
measures the number of chiral fermion zero modes minus
that of anti-chiral ones as follows from
the chiral rotation, (\ref{anomalyp}) is equivalent to the index theorem in mathematics.

Topological terms for $d$ = even can be derived from the
topological terms in $d+1$ dimensions considered above:
They are all Chern-Simons type
and obtained as a boundary contribution
from a $(d+1)$-dimensional topological term.
%follow from the $d$ = odd case by surface integral.

One could check that the topological response derived from
chiral anomaly is fully consistent with the periodic table;
(i) the topological terms derived in this way preserve
correct discrete symmetries for primary series;
(ii) for symmetry classes which are realized as topological
SC (i.e., no charge conservation), the anomaly polynomial
predicts that there is no topological response for EM field;
only gravitational (thermal) response exists.

One could think of this as an alternative ``derivation'' of
the periodic table (for the primary series).

\subsection{topological phases and D-branes}
\label{topological phases and D-branes}

\paragraph{introduction}

We would now like to point out an interesting
connection between topological phases in condensed matter
and D-branes.
D-branes are topologically stable objects in string theory;
At the level of classical (super)gravity, which is a low-energy
effective field theory of string theory,
D-brane can be visualized as a $(p+1)$-dimensional solitonic solution
to the 10d Einstein equation
(such D-brane is called D$p$-brane).
Besides such geometrical attribute in gravity theory,
an important property of D-branes for our purposes is the fact that
open string excitations on D-branes
give rise to a gauge field theory.
This dual nature of D-branes
has been proved to be a useful tool to understand
non-perturbative phenomena in gauge field theories,
such as monopoles, Seiberg-Witten theory
\cite{MQCD}, 
etc.
(Such dual nature of D-branes also plays a key role
in AdS/CFT or holography.
For our discussions in this article, however,
we will not use AdS/CFT.)

We will describe below,
what D-branes are,
how we can construct topological phases from D-branes,
and
why they give us an insight on the periodic table.
In fact, the opposite is also true;
topological phases deepen our understanding of the stability of D-branes.
As we will argue below, the stability of the so-called
``non-BPS'' D-branes is directly related to
the $\mathbb{Z}_2$ topological index of topological
insulators.

\paragraph{K-theory charge of D-branes}

\begin{table}
\begin{center}
\begin{tabular}{cccccccccccc}\hline
    & D($-1$) & D0 & D1 & D2 & D3 & D4 & D5 & D6 & D7 & D8 & D9 \\ \hline\hline
  type IIB & $\mathbb{Z}$ & 0 & $\mathbb{Z}$ & 0 & $\mathbb{Z}$ & 0 & $\mathbb{Z}$ & 0 & $\mathbb{Z}$ & 0 & $\mathbb{Z}$ \\
   type IIA & 0 & $\mathbb{Z}$ & 0 & $\mathbb{Z}$ & 0 & $\mathbb{Z}$ & 0 & $\mathbb{Z}$ & 0 & $\mathbb{Z}$ & 0  \\
    O9$^{-}$ (type I) & $\mathbb{Z}_2$ & $\mathbb{Z}_2$ & $\mathbb{Z}$ & 0 & 0 & 0 & $\mathbb{Z}$ & 0 & $\mathbb{Z}_2$ & $\mathbb{Z}_2$ & $\mathbb{Z}$
 \\ \hline
\end{tabular}
\end{center}
\caption{\label{dbrane}
D$p$-brane charges from K-theory, classified by
$\mathrm{K}(\mathbb{S}^{9-p})$,
$\mathrm{K}^{-1}(\mathbb{S}^{9-p})$ and
$\mathrm{KO}(\mathbb{S}^{9-p})$.
A $\mathbb{Z}_2$ charged D$p$-brane with
$p$ even or $p$ odd represents
a non-BPS D$p$-brane
or a bound state of a D$p$ and an anti-D$p$ brane, respectively.
}
\end{table}

As mentioned above,
D-branes are a topologically stable object in sting theory;
D$p$-brane is a $(p+1)$-d solitonic object
in 10d space-time of classical (super)gravity theory.
What is behind the stability of D-branes is the fact
that there is a ``charge'' associated to them.
These charges are quantized, and hence they cannot change for a smooth
deformation of field configurations;
D-branes are thus stable.

The charge of a stable D-brane can be either integer-
or $\mathbb{Z}_2$-valued, depending on the types of D-branes
and sting theory.
This is summarized in Table \ref{dbrane}.
Observe that for type IIA and IIB string theory,
``0'' and ``$\mathbb{Z}$'' appear in an alternating fashion.
On the other hand, for type I sting theory,
the way K-theory charges ``0'', ``$\mathbb{Z}$'' and ``$\mathbb{Z}_2$''
appear closely follows the Bott periodicity.
Compare 
entries ``type IIB'' and ``type IIA'' in 
Table \ref{dbrane}
with ``complex'' symmetry classes (A and AIII) in Table \ref{periodic table with gravity},
and entries ``type I'' in
Table \ref{dbrane}
with eight ``real'' symmetry classes in Table \ref{periodic table with gravity}.
In fact, it was argued that
D-brane charges are classified by K-theory
\cite{MiMo,WiK,HoK};
one then cannot help speculating on a possible connection between
D-branes and the periodic table of topological phases.

For some cases where a D-brane has an integral charge,
D-branes are an electric or a magnetic source of
an Abelian $p$-form gauge field
$C_{(p)} =
(1/p!)
C_{(p)}^{\mu_1 \mu_2 \cdots \mu_{p}}dx_{\mu_1}dx_{\mu_2} \cdots dx_{\mu_{p}}$,
the so-called ``Ramond-Ramond'' (RR) gauge field.
An integral of the RR-gauge flux generated by
a magnetic D-brane,
$\int_{S} dC_{(p)}$,
measures the charge of the D-brane,
where
$S$ is a hypersurface which encloses the D-brane.
For D-branes with a $\mathbb{Z}_2$ charge,
while we do see K-theory charge exist,
it is not possible to write it down
as a quantized integral of the RR flux.

\paragraph{D-brane configurations for topological phases}

\begin{figure}[tb]
\begin{center}
    \includegraphics[height=.3\textwidth]{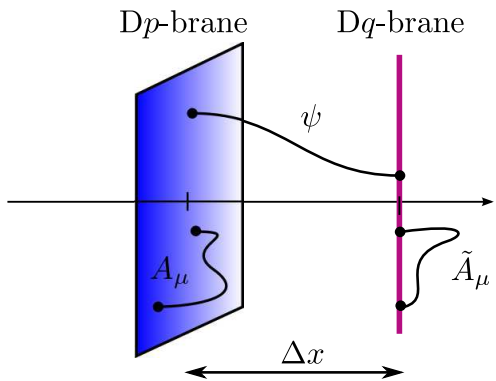}
\caption{
A typical D-brane configurations that realizes
a topological insulator (superconductor).
}
\label{brane}
\end{center}
\end{figure}

For our purposes, we consider D-brane configurations
which consist of two types of D-branes,
a D$p$-brane
and
a D$q$-brane.
They are located in parallel
and do not intersect
(Fig.\ \ref{brane})
\cite{DbranesTI,DbranesTIPRD,comment1}.
One can then ask what kind of (gauge) field theory is
realized in such D-brane configuration.
In string theory, quantum fields are realized as a
vibration of a fundamental string.

In our configuration,
an open string can have its end points on
D-branes.
Let us first consider a string which has one end on
a D$p$-brane and the other on a D$q$-brane.
Analyzing the vibrations of such string,
one finds a massive fermion in the string spectrum.
The mass of the fermion is proportional to the distance
between the D-branes.
For a relativistic fermion which is fully gapped by a mass term,
following the periodic table of topological phases,
one can discuss (under a suitable set of discrete symmetries)
its topological stability against adiabatic deformation;
one can ``compute'' its topological invariant or topological charge.
On the other hand, for the D-brane configuration at hand,
if it is stable, one can assign a topological (K-theory) charge to it,
following the K-theory charge of D-branes.
One can check these two types of topological charges agree precisely.
This is the first implication of the correspondence between
the periodic table and D-branes.

The Dirac fermion is not the only quantum field realized
in the D-brane configuration.
Let us now turn our attention to a string which has
its both ends on the D$p$-brane.
Such string vibration gives rise to a gauge field $A_{\mu}$
living on the D-brane.
(Its gauge group depends on the type of sting theory and D-branes.)
What is the dynamics of such gauge field on the D-brane?
This is answered by the effective action of D-branes.
The topological part (the so-called ``Wess-Zumino term'')
of the effective action for a D$p$-brane in a flat space is%
\footnote{
For the discussion below, we will focus on the situation where
the topological charge of D-branes is an integer, for presentational
simplicity.
}
\begin{equation}
S_{\mathrm{WZ}} =
\sum_q
\int_X
C_{(q+1)}
\wedge
\mathrm{ch}(E)
\hat{A}(TX)
\label{RRC simple}
\end{equation}
where integration ($\int_X$) is over the $(p+1)$-d 
world-volume $X$ of the D$p$-brane.%
\footnote{
This is obtained from a more general expression (\ref{RRC})
by noting that
in our situation $TX\otimes NX$ is a trivial bundle,
and hence
$\hat{A}(TX\otimes NX)
=
\hat{A}(TX)\wedge\hat{A}(NX)
=1$.
}
In this action,
$C_{(q+1)}$ is the background $(q+1)$-form RR gauge field
(which is, in our situation,  sourced by the D$q$-brane).
On the other hand,
$\mathrm{ch}(E)$ and $\hat{A}(TX)$
depends on the field configuration on the D$p$-brane;
The U(1) gauge bundle on the brane is denoted by $E$;
$TX$ and $NX$ are the tangent bundle and normal bundle.
Plugging the RR-field generated by the D$q$-brane,
whose integral $\int dC_{(q+1)}$
is quantized as it measures the topological charge of
the D$q$-brane,
the Wess-Zumino term of the D$p$-brane
recovers precisely
the response theory that we discussed in terms of anomaly.

As an example, let us consider the case with $p=q=5$;%
\footnote{
Since we can apply
the T-duality equivalence which shifts the value of $p$ and $q$ by one 
\cite{PolT}, we can fix the values of
$p$, say $p=5$.
}
\footnote{
This configuration has 6 ND directions.
}
this D-brane configuration realizes the 2d QHE.
The vibration of an open sting stretching between the D-branes
gives rise to a (2+1)d massive Dirac fermion,
$
\mathcal{H}
=
\psi^{\dag} (-{i} \sum_{i=x,y} \sigma_i \partial_i + m \sigma_z)\psi
$
where $\psi$ is the two-component complex Dirac fermion field.
On the other hand, the WZ action of D$p$-brane is given by
the Chern-Simons term:
\begin{align}
S_{\mathrm{WZ}}
&=
\frac{1}{2 (2\pi)^2}
\int_X
C_{(2)}
\wedge
F
\wedge
F
=
\frac{1}{2 (2\pi)^2}
\int_{Y\times Z}
C_{(2)}
\wedge
d (A\wedge F)
\nonumber \\%%%%%
&=
\frac{-1}{8 \pi^2}
\int_Z
dC_{(2)}
\int_Y
A\wedge F
=
\frac{\pm m}{8\pi |m|}
\int_Y
A \wedge dA.
\end{align}
Here,
the integral is over
the 6d world volume of the
D$p$-brane,
which we split into the common directions of the D$p$- and D$q$-branes
($Y$)
and the compliment thereof ($Z$),
and we noted that
the D$q$-brane couples magnetically
to the RR two form $C_{(2)}$, and hence
the integral $(2\pi)^{-1} \int dC_{(2)} = \pm 1/2$ measures
the RR-charge of the D$q$-brane.
(We have done a partial integration.)
The sign $\pm$ in front of the Chern-Simons term
corresponds to 
$\mathrm{sgn}\, \Delta x =\pm 1$
(see Fig.~\ref{brane}).
Similarly,
from the WZ coupling
$
S_{\mathrm{WZ}}
=
\int_X
C_{(2)}
\wedge
\hat{A}(R)
=
\frac{1}{12}
\int_X
C_{(2)}
\wedge
\frac{ {i} R}{4\pi}
\wedge
\frac{ {i} R}{4\pi}
$,
we obtain
the gravitational Chern-Simons term
$\sim \omega\wedge d\omega+\frac{2}{3}\omega^3$,
as expected from the responses.

Observe the structural parallelism
between
the topological terms in the response theory
of topological phases,
and the Wess-Zumino action of D-branes.
In the former, the effective action looks, typically, as
$I_{\mathrm{eff}} \propto (\mbox{topological invariant})$
$\times
\int d^{d+1}x\, (\mbox{topological term in gauge theory})$,
where
``(topological invariant)'' is the topological invariant
of the topological phase in question,
and
``(topological term in gauge theory)''
is the term of topological origin in gauge theories such as
the Chern-Simons term, or the axion term.
For example, 
$I_{\mathrm{eff}} \propto \mathrm{Ch}
\times
\int d^3x\, \epsilon^{\mu\nu\rho} A_{\mu} \partial_{\nu} A_{\rho}$
in the QHE, where
$\mathrm{Ch}$ is the TKNN integer.
In the D-brane construction of topological phases,
the coefficient in front of the topological term
is given by the integral of the RR-field,
and measures the K-theory charge of the D$q$-brane
\cite{commentrealsymmetry,commentedge}.

\section{conclusion}

We have described,
in the order of increasing spatial dimensions,
from two, three, and to arbitrary dimensions, 
the theory of thermal response in topological 
phases.
Emphasized is a close analogy to 
electromagnetic responses of topological insulators
by adopting the language of gravitomagnetism,
which revealed a cross-correlation between 
thermal and mechanical (rotational) responses.

Let us close with discussion on effects of interactions.
The fact that topological currents
(either in electromagnetic or thermal response)
are related to anomalies in field theories 
suggests topological phases with topological currents
should be stable against interactions. 
This can be seen simply by observing that
a term of topological origin in the effective action,
once exists, has its coefficient which is quantized
(in some case in the presence of some discrete symmetry). 
Thus, small interactions should not destroy 
topological properties of a given topological phase.%
\footnote{
An alternative derivation of the 
electrical/thermal Str\v{e}da formula 
mentioned in Footnote \ref{footnoteCFT}
also provides another link between
topological currents
and 
anomalies,
suggesting the stability of topological 
currents against interactions. 
}

We have computed the theta 
term (both electromagnetic and gravitational ones) 
by making use of chiral anomaly
(which arises as a Fujikawa Jacobian). 
Let us imagine repeating the same calculations in the presence
of interactions such as 
$\varphi \bar{\psi}\psi$
(a coupling to a bosonic scalar field $\varphi$), 
or electromagnetic interactions. 
The Adler-Bardeen theorem says that the 
anomaly (the Fujikawa Jacobian)
will not be altered even in the presence of such interactions. 
(Historically,
this nonrenomalization of the chiral anomaly
was important to predict the number of quark colors
even at the time when the details of the strong interaction were not known. )

Insensitivity of anomalies to interactions can further be illustrated
by the ``anomaly matching condition" proposed by 't Hooft; 
an anomaly (i.e., topological current) can be computed 
in terms of either the infra-red (IR) or ultra-violet (UV) 
degrees of freedom, and the results should be the same. 
For example, 
the degrees of freedom in solids at UV are of ``free-fermion" type, 
while deep in the IR region, 
such description can be replaced by quasi-particles 
(such as excitons) that arise due to interactions.

The discussion above naturally echoes
in the D-brane construction of topological phases;
the field theories realized by D$p$-D$q$ systems 
come with, in addition to massive fermions we discussed, 
other fields and interactions among them. 
As we can understand more or less geometrically, 
the topological phases are nevertheless stable.

While we have discussed the stability of 
non-interacting topological phases against 
interactions, it remains largely an open problem 
what is the nature of
strongly interacting topological phases 
which arise solely because of interactions,
if they exist at all
beyond the fractional quantum Hall effect.  
The existence of a topological current,
however, should be a hallmark
that we can use to characterize 
even for these putative ``fractional topological insulators''.

\section*{Acknowledgements}

We thank 
Taylor Hughes, 
Shunji Matsuura,
Joel Moore,
Xiao-Liang Qi,
Mike Stone, 
Ashvin Vishwanath, 
and
Shou-Cheng Zhang 
for useful discussions.
This work is
supported by MEXT Grand-in-Aid No. 20740167,
19048008, 19048015, 21244053, Strategic International
Cooperative Program (Joint Research Type) from Japan
Science and Technology Agency, and by the Japan Society
for the Promotion of Science (JSPS) through its ``Funding
Program for World-Leading Innovative R$\&$D on Science
and Technology (FIRST Program).''


\appendix

\section{a short course for D-branes}

\subsection{what is D-brane?}

In string theory
\cite{PolT}, 
we consider a string as a fundamental object
instead of an elementary particle.
The vibration of a closed string  
produces, among others, the gravitational field at low energies. 
On the other hand,
the vibration of an open string,
which looks topologically like an interval with 
two ends,
produces, among others, 
gauge fields in the low-energy limit.

A D-brane is defined as an object where open strings can end on it
\cite{PolT,PolD};
D-brane can be considered as an object giving a boundary condition
to string vibrations -- the name D-brane originates from
``Dilchlet'' brane.

In terms of the 10d gravity theory, 
which is the low-energy limit of string theory,
D-branes are a stable solitonic object which is quite massive. 
%(its stability actually should not be limited to low energy).
A D$p$-brane extends in $p$-spatial directions and in time-direction;
i.e, its world volume is $(p+1)$-d.
For example,
a D0-brane and a D1-brane 
look like a particle (called D-particle) and
a string (called D-string),
respectively. 

For our purposes, we will note the following properties
(see Ref.\ \cite{PolD}):
(i)
On a D-brane, an abelian gauge theory is realized.
The fluctuations of the gauge field correspond to
the fluctuations of the end points of open string.
If we have $N$ coincident D$p$-branes,
we obtain a non-abelian $\mathrm{U}(N)$ gauge theory.
(ii)
An intersection of two D-branes realizes massless fermions.
(iii)
D-branes are characterized by a K-theory charge. 
The last point will further be discussed below.

\subsection{K-theory classification of stable D-branes}

Gravitons and gauge fields are not the only fields 
that arise in superstring theory. 
In addition to them, 
the vibration of a closed string generates
Abelian $p$-form gauge fields, 
the Ramond-Ramond (RR) gauge fields. 
(Here, we are focusing for a moment on 
a particular type of superstring theory,
type IIA and IIB superstring theory (with or without orientifolds). 
Supersymmetric (or BPS) D$p$-branes are charged under 
these gauge fields;
they have the RR charges, and hence stable.
A D$p$-brane directly (electrically) 
couples to the $(p+1)$-form RR field $C_{(p+1)}$.
An anti D$p$-brane is defined to be the one with a negative RR charge.
In type IIA superstring,
$p$ takes only even integer values i.e., $p=0,2,4,6,8$ and
in type IIB, $p$ takes only odd ones $p=-1,1,3,5,7,9$.

Moreover, a D$p$-brane couples to other RR $q$-forms with $q<p$
in the presence of the gauge flux on the brane.
This is clearly described by the following formula
of the RR couplings of a D$p$-brane
\cite{GHM}:
\begin{equation}\label{RRC}
S_{\mathrm{RR}} =
\sum_q
\int_X
C_{(q+1)}
\wedge
\mathrm{ch}(E_D)
\sqrt{\frac{\hat{A}(TX)}{\hat{A}(NX)}},
\end{equation}
where $TX$ and $NX$ are the tangent bundle
(i.e., the vector bundle tangent to the
D-brane world-volume $X$) and normal bundle
(i.e., the vector bundle which is normal to $TX$).
The U(1) gauge bundle on the brane is denoted by $E_D$.
This formula (\ref{RRC}) means that if there is a non-trivial gauge bundle
on the D$p$-brane, it is possible that there exist charges
which correspond to lower dimensional D-brane charges.
Such a configuration can be interpreted as a bound state
of a D$p$-brane and the lower dimensional D-branes.

In fact, there are D-branes which do not have any RR charges
and which are nevertheless stable.
They are not supersymmetric and
are called non-BPS D-branes \cite{BeGa}. 
Also a system of a D$p$-brane and an
anti D$p$-brane sometimes forms a stable bound state.
Such a system is called
a brane anti-brane system
\cite{Sen}.
They typically exist
in the presence of the special projection called the
orientifold projection.
These brane configurations exhaust all possible
D-branes in string theory.
The orientifold projection means that we require the invariance of
string theory under the action
$\tilde{\Omega_q}=I_q\cdot \Omega$,
defined by the product of the parity $I$
with respect to $q$ spacial coordinates
and the orientation reverse $\Omega$ of the string world-sheet.
The set of fixed points of $I_q$ is called the orientifold $(9-q)$ plane.

Being a stable object in string theory, one could imagine what is
protecting them to ``decay''. 
It turns out that one can assign a K-theory charge to
D-branes,
which is the reason of stability. 
Indeed, the K-theory provides a very systematic classification of D-branes
in string theory \cite{WiK,HoK}. 
In mathematics, (topological)
K-theory classifies vector bundles. More precisely, we start from a pair of
two bundles $(E_1,E_2)$ on a manifold $X$ and consider its difference.
In other words, we
introduce the identification
\begin{equation}\label{Kthi}
(E_1\oplus H,E_2\oplus H)\simeq (E_1,E_2).
\end{equation}
This defines the K-group $K(X)$.

In string theory, this identification is naturally interpreted as follows.
We start with a
brane anti-brane system.
The gauge bundles on the D-brane and the anti D-brane are regarded
as $E_1$ and $E_2$.
Typically a brane anti-brane system becomes unstable
because the total RR charge is vanishing and it can pair-annihilate.
In other words, there appears so-called
a tachyon field in the open string between the brane and the anti-brane. 
The tachyon field has unstable potential energy 
and condenses like the Higgs effect, 
which makes the system decay into a 
lower-dimensional D-brane.
This procedure is mathematically described by (\ref{Kthi}),
which means that the charge is conserved under the tachyon condensation.
If the gauge field configurations are the same i.e., $E_1=E_2$,
then the brane and anti-brane are completely annihilated, and
nothing remains after the tachyon condensation.
If $E_1\neq E_2$, then eventually the system decays into a D-brane
which corresponds to the difference between $E_1$ and $E_2$.
We presented the K-theory classification of D-branes in type IIA,
type IIB and type I string theory, where we take $X=\mathbb{S}^{9-p}$
for a D$p$-brane via a compactification procedure.
Notice that type I string theory is defined to be the projection of
type IIB string theory by an (SO type) orientifold 9-plane.
For type IIA and type I, we need to use a different K-theory
called $K^{-1}(X)$, which just shifts the dimension by one,
and $KO(X)$, which is the real valued version of $K(X)$.

In particular, if we ignore the torsion of K-group $K(X)$,
then it is known that $K(X)$ is reduced to 
even-dimensional
cohomology $\oplus_{i\geq 0}H^{2i}(X,\mathbb{Q})$. 
Indeed, this is explicitly given by the Chern character 
and this nicely matches with the RR coupling formula (\ref{RRC})
\cite{MiMo}. 
On the other hand, the argument based on K-theory with torsion
is quite general and includes the case where the D-branes do not have
any RR charges as is so for the non-BPS D-brane.


\begin{thebibliography}{00}
\bibitem{TSF}
$^3$He-B should be called topological superfluid,
rather than topological SC. 
Since we are interested in topological properties of 
fermionic quasiparticle wavefunctions, 
and since thermal responses we are after are generic
and common to both of them,
we do not distinguish topological superfluid 
and topological SCs in this paper.
\bibitem{comment1}

D$p$-D$q$ systems have common $(d+1)$ dimensions;
this is where a $d$-dimensional topological insulator exists.
From the viewpoint of open string,
the $(d+1)$ dimensions has the Neumann boundary condition at the both ends
(called the NN direction).
In the same way, $(p+q-2d)$ coordinates are along Neumann-Dirichlet (ND)
directions, while
the remaining $(9-p-q+d)$ ones are in the DD direction.
\bibitem{commentrealsymmetry}

In this section, for illustration purposes, we have not discussed 
how discrete symmetries are implemented in the D-brane realization of 
topological phases. 
We found that particle-hole symmetry corresponds to 
the orientation reverse $\Omega$ of strings,
while
chiral (or sublattice) symmetry to
the parity transformation in one of the DD directions, respectively. 
Time-reversal is 
nothing but the orientifold projection.

With these implementations of discrete symmetries, 
our D-brane systems are characterized by the directions of
D$p$, D$q$ and the orientifold.
Classifying these conditions 
(aided by an analysis on the stability of boundary modes of these systems), 
we found that the allowed D-brane systems are
in one-to-one correspondence to the ten-fold classification of 
topological phases.
\bibitem{commentedge}
Their edge states are obtained by replacing one of DD directions 
by a ND direction.
The chiral fermions typical in edge states appear due to 
the topological mechanism.  
This construction leads to the correct fermion spectra 
for the topological insulators
and their edge states.
\end{thebibliography}
\end{document}